\documentclass[aps,float,prd,psfig,amsmath,twocolumn]{revtex4}
\usepackage[dvipdfmx]{graphicx}
\usepackage{amsmath,amssymb,natbib,siunitx,booktabs,url}
\usepackage{color}
\newcommand{\add}[1]{\textcolor{red}{#1}}

\newcommand{\vect}[1]{\mbox{\boldmath${#1}$}}
\newcommand{\lmk}{\left(}
\newcommand{\rmk}{\right)}

\newcommand{\beq}{\begin{equation}}
\newcommand{\beqa}{\begin{eqnarray}}
		  \newcommand{\eeq}{\end{equation}}
\newcommand{\eeqa}{\end{eqnarray}}

\begin{document}

\title{A Multimessenger Strategy for Downselecting the Orientations of Galactic Close White Dwarf Binaries}

\author{Naoki Seto }
%%%%%%%%%%%%%%%%%%%%%%%%%%%%%%%%%%%%%%%%%%%%%%%%%%%%%%%%%%%%%%%%%%%%%%%%%%
\affiliation{Department of Physics, Kyoto University, 
Kyoto 606-8502, Japan
}

\date{\today}

\begin{abstract}
The planned space-based gravitational wave detector, LISA, will provide a fundamentally new means of studying the orbital alignment of close white dwarf binaries. However, due to the inherent symmetry of their gravitational wave signals, a fourfold degeneracy arises in the transverse projections of their angular momentum vectors. In this paper, we demonstrate that by incorporating timing information from electromagnetic observations, such as radial velocity modulations and light curves, this degeneracy can be reduced to twofold. 

\end{abstract}


\maketitle

\section{introduction}
Do the orbital angular momenta of  Galactic binaries statistically  align with Galactic structures?  For over 100 years, astrometric and radial velocity data have served as fundamental observational resources for investigating this question (see, e.g., \cite{1929AJ.....40...11C} for early studies). In a recent detailed study,  Agati et al. \cite{2015A&A...574A...6A} analyzed  95 binaries within 18 pc of the Sun and found that their angular momenta are consistent with being randomly oriented.

More recently, in 2023, Tan et al. \cite{2023ApJ...951L..44T} examined the orientations of the symmetry axes of 14 planetary nebulae that host (or are inferred to host) short-period binaries around the Galactic bulge. They reported that the axes are not randomly oriented but tend to be parallel to the Galactic plane. Planetary nebulae consist of gas expelled during the formation of white dwarfs, and the observed 14 axes are considered to coincide with the orbital axes of their associated binaries. Tan et al. \cite{2023ApJ...951L..44T} proposed that the Galactic magnetic field at the time of binary formation may be responsible for this observed anisotropy. This finding contrasts sharply with the results of Agati et al. \cite{2015A&A...574A...6A}, who analyzed local binaries, highlighting the need for further observational studies on this issue.

The Laser Interferometer Space Antenna (LISA) is designed to have optimal sensitivity to gravitational waves (GWs) in the 0.1-10 mHz range and is anticipated to detect approximately $10^4$ Galactic close white dwarf binaries (CWDBs) emitting nearly monochromatic GWs \cite{amaro2023astrophysics} (see also \cite{hu2017taiji} for Taiji and \cite{luo2016tianqin} for TianQin). 
{At relatively high frequencies above $\sim3$ mHz, CWDBs will likely be individually resolved throughout the Galaxy \cite{amaro2023astrophysics}. In contrast, at lower frequencies, the number of CWDBs per frequency bin is expected to be much greater than unity, with only a small fraction near the Sun being resolvable. The unresolved sources contribute to the confusion foreground noise \cite{amaro2023astrophysics}.}

Due to dissipative effects, most of these CWDBs are expected to have nearly circular orbits. Since both the generation and measurement of GWs are inherently geometrical, LISA will provide a fundamentally new means of studying the orbital orientations of Galactic binaries \cite{Seto:2024odc,Seto:2024wrw}.

Unfortunately, due to the intrinsic symmetry of their gravitational waveforms, there is a fourfold degeneracy in estimating the transverse projections of orbital orientations, a well-known issue in the GW community (see, e.g., \cite{Cornish:2003vj} in the context of Galactic binaries).  For instance, suppose LISA detects an edge-on CWDB located near the Galactic plane, with its angular momentum aligned parallel to the plane. The fourfold degeneracy would prevent us from determining whether the binary's orientation is truly parallel or instead perpendicular to the Galactic plane. Consequently, relying solely on LISA data imposes significant limitations on our ability to analyze the orientations of CWDBs \cite{Seto:2024odc,Seto:2024wrw}.

Meanwhile, CWDBs emit electromagnetic (EM) signals in addition to GWs \cite{amaro2023astrophysics}. In the LISA era, CWDBs will be key observational targets for multimessenger astronomy.
Their sky positions  can be estimated from the amplitude and Doppler modulations induced by the motion of the GW detectors \cite{Cutler:1997ta}. {
At GW frequencies above $f_{\rm gw} \sim 1$ mHz, Doppler modulation generally becomes more useful. For an observational time longer than 2 years, the typical size of the error ellipsoid in the sky is estimated to be $\sim 0.5\, {\rm deg}^2 (\rho/50)^{-2} (f_{\rm gw}/5\,{\rm mHz})^{-2}$, where $\rho$ is the signal-to-noise ratio \cite{Takahashi:2002ky}.}

 The light curves and the radial velocity data of CWDBs are modulated by orbital motion and should be relatively easy to observe (see, e.g., \cite{Burdge:2019hgl}). In particular, nearly edge-on CWDBs will exhibit distinct eclipsing patterns, making them ideal candidates for follow-up EM observations. Indeed, according to Korol et al. \cite{Korol:2017qcx}, as many as $\sim$100 eclipsing CWDBs ({nearly at edge-on configurations}) could be observed simultaneously by LISA and the Vera C. Rubin Observatory.  {   
 The majority of these $\sim 100$ CWDBs will be at distances less than $\sim 4$ kpc and thus belong to the disk component, comprising a small fraction of the $\sim 10^4$ resolved CWDBs (see also \cite{Littenberg:2012vs}).}

In this paper, we explore the possibility of reducing the fourfold degeneracy of the angular momentum vector, by leveraging multimessenger observations of CWDBs. We revisit the time profile of their GW signals, incorporating orbital phase information partially inferred from basic EM data. We then propose a simple method to reduce the fourfold degeneracy  to twofold. {
In our explanation of the fourfold degeneracy, we considered the example of a hypothetical edge-on binary system located near the Galactic plane, with its angular momentum aligned parallel to the plane. Our reduction method enables us to eliminate the two spurious orientation solutions that are perpendicular to the plane.}
The approach presented in this paper could further motivate deeper follow-up searches for CWDBs detected by LISA, extending beyond those exhibiting eclipsing patterns.
\if0
 \add{targeting EM counterparts beyond the eclipsing systems. Our observational resources there would be the Doppler modulation and the light curve modulation e.g., induced by  the ellipsoidal variation and irradiation effects \cite{Burdge:2019hgl}.    }
 \fi

\section{GW from a circular binary}
In this section, we review the GW emission from a circular binary and discuss how to estimate its orbital orientation based on the observed GW signal.
\subsection{Orbital motion}
Let us consider a circular binary consisting of two stars, $\alpha$ and $\beta$.  
We denote its orbital separation by $a$ and the masses of the stars by $m_\alpha$ and $m_\beta$, respectively.  
The orbital frequency is given by  
\beq  
f = \frac{1}{2\pi} \lmk \frac{G M_T}{a^3} \rmk^{1/2}  
\eeq  
where the total mass is $M_T = m_\alpha + m_\beta$.

We introduce the coordinate system $XYZ$, as shown in Fig. 1.  
The binary orbits in the $XY$-plane (around the origin), and the orientation vector ${\vec e}_j$ of its angular momentum is directed along the $Z$-axis.  
The positions of the two masses are given by  
\beqa
(x_\alpha,y_\alpha,z_\alpha) &=& a\frac{m_\beta}{M_T} (\cos\Phi_{\rm s}(t),\sin\Phi_{\rm s}(t),0), \label{xa}\\
(x_\beta,y_\beta,z_\beta) &=& -a\frac{m_\alpha}{M_T} (\cos\Phi_{\rm s}(t),\sin\Phi_{\rm s}(t),0), \label{xb}
\eeqa
where the orbital phase of the binary is  
\beq
\Phi_{\rm s}(t) = 2\pi f t + \varphi_{\rm s}.
\eeq

\subsection{Observed GW}
Next we discuss  GW emission from the binary. Since CWDBs have small post-Newtonian parameters $O(10^{-3})$ in the LISA band, we can apply the quadrupole formula (with the exception of possible rare cases \cite{seto25}).

\begin{figure}[t]
 \includegraphics[width=0.9\linewidth]{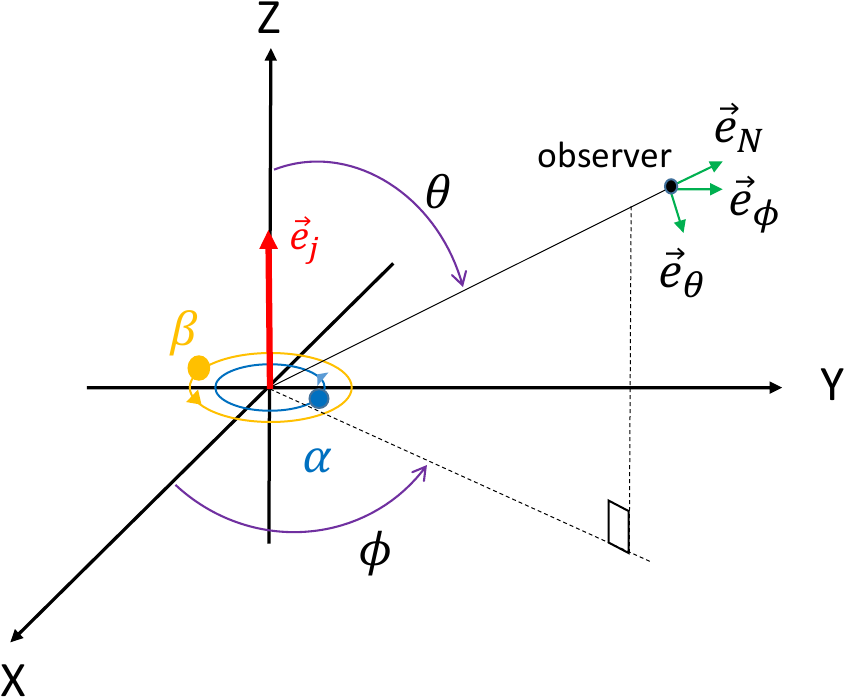} 
 \caption{Configuration of a binary and an observer.  The binary  rotates in the $XY$-plane with its orientation vector ${\vec e}_j$ pointing toward the $Z$-axis. At the observer, we introduce the orthonormal vectors $({\vec e}_N,{\vec e}_\theta,{\vec e}_\phi)$.  
 }  \label{fig:con}
\end{figure}

We consider an observer at $(r,\theta,\phi)$ in the spherical coordinate system (see Fig. 1).  The polar angle $\theta$ corresponds to the inclination angle. We have $\theta=\pi/2$ for the edge-on and $\theta=\{0,\pi\}$ for the face-on configurations.  The two transverse vectors $({\vec e}_\theta,{\vec e}_\phi)$ compose the wave's principle axes (see e.g., \cite{Cutler:1997ta}). {Later, we virtually rotate the binary around the binary-observer axis, but continue to fix the reference directions $({\vec e}_\theta,{\vec e}_\phi)$ as shown in Fig. 1.  }

 As presented in various textbooks (see, e.g., \cite{2017mcpo.book.....T}), the GW signal at the observer is expressed as 
\beq
h_{\mu\nu}=h_+(t) e_{\mu\nu}^+ + h_\times(t) e_{\mu\nu}^\times
\eeq
with the transverse traceless polarization tensors
\beq
{\vect e}^+={\vec e}_{\theta}\otimes{\vec e}_{\theta}-{\vec e}_{\phi}\otimes{\vec e}_{\phi},~
{\vect e}^\times={\vec e}_{\theta}\otimes{\vec e}_{\phi}+{\vec e}_{\phi}\otimes{\vec e}_{\theta} \label{ptt}
\eeq
or equivalently
\beq
h_{\theta\theta}=-h_{\phi\phi}=h_+,~~h_{\theta\phi}=h_{\phi\theta}=h_\times.
\eeq 

The time-dependent functions $h_+(t)$ and $h_\times(t)$ are given by 
\beqa
h_+(t)&=&-A(\cos^2 \theta+1)\cos2\Phi_{\rm o}(t) \label{hp}\\
h_\times(t) &=&-2A\cos \theta\sin2\Phi_{\rm o}(t)\label{hc}
\eeqa
with the phase $\Phi_{\rm o}(t)$ and the amplitude $A$ at the observer  
\beqa
\Phi_{\rm o} (t)&=&2\pi f (t-r/c) +\varphi_{\rm s}-\phi\\
A&=&\frac{2G^{5/3}m_\alpha m_\beta M_T^{-1/3}(2\pi f)^{2/3}}{r}. \label{vp}
\eeqa
In Fig. 2, the black and gray curves show the typical shapes of the two functions $h_+ = h_{\theta\theta}$ and $h_\times = h_{\theta\phi}$ for $\theta = 31^\circ$. In the lower part of the figure, we also show the deformation pattern of the GW at representative epochs. The GW frequency is twice the orbital frequency $f$. 

\subsection{Fourfold degeneracy}

In this section, we discuss the estimation of the orientation vector ${\vec e}_j$ from the observed GW signal. Following conventions, we decompose the vector ${\vec e}_j$ into the inclination angle $\theta$ and the unit projection vector onto the transverse plane ($-{\vec e}_\theta$ in the case of Fig. 1). For the latter, the so-called polarization angle $\psi$ is often used, which is measured counterclockwise from a reference direction on the plane.

The inclination angle $\theta$ can be observationally determined within the range $[0, \pi]$ from the two orthogonal polarization coefficients $h_+$ and $h_\times$. More specifically, we use their relative amplitudes and phase (which corresponds to $\pm 1/4$ of the GW cycle). At the face-on configuration ($\theta = 0$ or $\pi$), the incoming GW becomes circularly polarized, and the orientation vector ${\vec e}_j$ can be uniquely determined, although its transverse projection is ill defined. Below, we exclude these two special configurations.

\if0
\add{
Next, in Fig. 1, let us virtually rotate the binary system and the associated GW around the binary-observer axis by \(90^\circ\). Under this rotation, the transverse-traceless tensor \(h_{\mu\nu}\) changes its sign due to its spin-2 nature. To demonstrate this, we introduce a pair of orthonormal transverse vectors, \(\vec{e}_1\) and \(\vec{e}_2\), fixed at the observer (for a given sky direction \(\vec{e}_N\)). Before the rotation, the principal axes can be expanded   as
\beq
{\vec e}_\theta= {\vec e}_1 \cos\eta+{\vec e}_2\sin\eta ,~~{\vec e}_\phi=- {\vec e}_1\sin\eta+ {\vec e}_2\cos\eta
\eeq
with an angular parameter  \(\eta\). After the \(90^\circ\) rotation, the angle becomes \(\eta + \pi/2\), and  the rotated vectors $({\vec e}_\theta\!', {\vec e}_\phi\!')$ are given by the original ones as  $({\vec e}_\theta\!', {\vec e}_\phi\!')=({\vec e}_\phi, -{\vec e}_\theta)$. We observe that the associated polarization tensors (\ref{ptt})  flip their signs.  }
\fi

{
Next, in Fig. 1, let us virtually rotate the binary and the associated GW around the binary-observer axis by $90^\circ$. Under this rotation, the transverse-traceless tensor $h_{\mu\nu}$ changes its sign, due to its spin-2 nature.  In terms of the original orthonormal vectors $\vec{e}_\theta$ and  $\vec{e}_\phi$ defined in Fig. 1, after the $90^\circ$ rotation, the transverse projection of the orbital orientation becomes \(-\vec{e}_\phi\). }

Importantly, from Eqs. (\ref{hp})-(\ref{vp}), this sign change can be absorbed into a phase shift $\varphi_{\rm s} - \phi \to \varphi_{\rm s} - \phi + \pi/2$.   Therefore, by adopting the new combination $(-{\vec e}_\phi, \varphi_{\rm s} - \phi + \pi/2)$ for the projected direction and phase, we can generate an observed GW signal identical to that obtained with the original combination $(-{\vec e}_\theta, \varphi_{\rm s} - \phi)$.  

Additionally, considering two other solutions corresponding to rotation angles of $180^\circ$ and $270^\circ$, we find a total of four possible candidates, $\pm {\vec e}_\theta$ and $\pm {\vec e}_\phi$ (in the original frame), for the transverse projection of ${\vec e}_j$. This fourfold degeneracy is well known in the context of parameter estimation with quadrupolar GWs \cite{Cornish:2003vj}.

\begin{figure}
 \includegraphics[width=1.\linewidth]{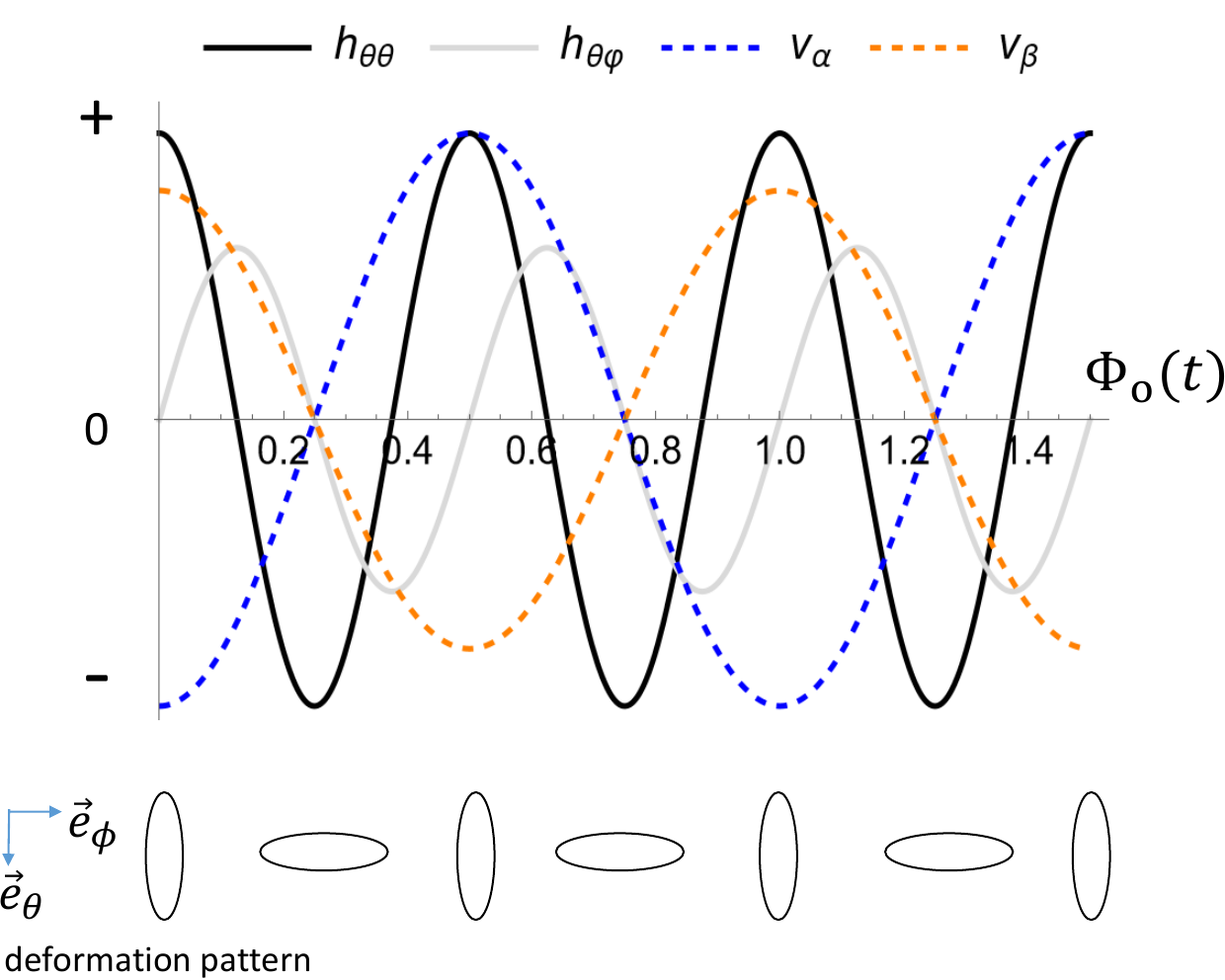} 
 \caption{Time profiles of the GW signal and the radial velocities. The specific values on the vertical axis are irrelevant to our study. The black curve represents the GW deformation pattern $h_+ = h_{\theta\theta} = -h_{\phi\phi}$, while the gray curve corresponds to $h_\times = h_{\theta\phi} = h_{\phi\theta}$. The dashed lines indicate the radial velocities of the two component stars, $\alpha$ and $\beta$. The ellipses at the bottom illustrate the tidal deformation patterns of the GW at representative epochs.
 }  \label{fig:del}
\end{figure}

\section{EM Signals}

In this section, we explain how to extract the orbital phase information of a CWDB using basic EM data, such as the time modulation of radial velocities and photometric luminosity. Note that these EM signals are invariant under rotation around the line of sight and, by themselves, do not provide azimuthal information.

\subsection{Doppler shifts}
We first evaluate the radial velocity components $v_\alpha$ and $v_\beta$ for the two stars, $\alpha$ and $\beta$. Taking the time derivatives of Eqs.~(\ref{xa}) and (\ref{xb}) and then computing their inner products with the radial unit vector $\vec{e}_N$, we obtain  

\begin{align}
    v_\alpha(t) &= -2\pi f a \frac{m_\beta}{m_T} \sin\theta \sin\Phi_{\rm o}(t), \label{v1} \\
    v_\beta(t) &= 2\pi f a \frac{m_\alpha}{m_T} \sin\theta \sin\Phi_{\rm o}(t). \label{v2}
\end{align}  

Here, we have ignored the trivial bulk velocity contribution. The velocities vanish, i.e., $v_\alpha = v_\beta = 0$, at $\sin \Phi_{\rm o}(t) = 0$. Note that the phase function $\Phi_{\rm o}(t)$ in Eqs.~(\ref{v1}) and (\ref{v2}) is identical to that appearing in Eqs.~(\ref{xa}) and (\ref{xb}) for the GW signal. Therefore, at $\sin \Phi_{\rm o}(t) = 0$, the strain deformation satisfies $h_{\theta\theta} = -h_{\phi\phi} < 0$ and $h_{\theta\phi} = h_{\phi\theta} = 0$, indicating compression along the $\pm \vec{e}_\theta$ directions (see Fig.~2).  

By comparing the radial velocity profiles with the strain pattern, we can refine the transverse projection of the orientation vector $\vec{e}_j$ to lie along the $\pm \vec{e}_\theta$ directions. This reduces the original fourfold degeneracy ($\pm \vec{e}_\theta$ and $\pm \vec{e}_\phi$) that arises when using only GW signals.  

Since we are dealing with four discrete options ($\pm \vec{e}_\theta$ and $\pm \vec{e}_\phi$), we do not need to determine the exact epochs at which $\sin \Phi_{\rm o}(t) = 0$ from the radial velocity data. A phase accuracy of $\Delta \Phi_{\rm o}(t) \sim 0.1$ would be sufficient for the present scheme.

\subsection{Light curve}
Next, we discuss a similar scheme based on the periodic variation of the light curve. We may observe signatures induced by strong binary interactions, such as ellipsoidal variations and irradiation effects \cite{2016A&A...591A.111M}. Here, for simplicity and ease of follow-up observations, we focus on the use of eclipse timing.

After straightforward calculations, we find that the transverse distance between the two stars is proportional to  
\[
[\sin^2\Phi_{\rm o}(t)\sin^2\theta + \cos^2\theta]^{1/2}.
\]  
For a nearly edge-on configuration with \(\theta \sim \pi/2\), sharp drops in the light curve can be observed around the conjunction epochs, where \(\sin\Phi_{\rm o}(t) = 0\) \cite{Burdge:2019hgl}. Therefore, as in the case of velocity curves, by examining the strain pattern during eclipses, we can identify the two candidate orientations \(\pm \vec{e}_\theta\), while excluding the spurious solutions \(\pm \vec{e}_\phi\).

\section{Summary}

LISA is expected to detect approximately $10^4$ CWDBs emitting nearly monochromatic GWs. By geometrically analyzing the waveform from each binary, we can obtain information on its orbital orientation vector, ${\vec e}_j$, in a fundamentally new way. However, due to the intrinsic symmetry of GW emission, there is a fourfold degeneracy in the projected direction of the vector ${\vec e}_j$ onto the transverse plane.

In this paper, by revisiting the time profile of the GW signal in response to orbital motion, we propose a multimessenger strategy to reduce this degeneracy to twofold. The key idea is to identify the strain deformation pattern at specific orbital phases, inferred from EM data such as radial velocity curves and light curves.

Eclipsing patterns in photometric data will serve as the primary observational signature for follow-up identification of CWDBs initially detected by LISA. To expand the sample of binaries available for orientational analysis, an extensive spectroscopic analysis for short-period binaries would be highly beneficial.

\if0
\acknowledgements
 This work is supported by JSPS Kakenhi Grant-in-Aid
for Scientific Research (Nos. 19K03870 and
23K03385).
\fi

%\input{ref.tex}

% Create the reference section using BibTeX:
\bibliography{ref}

\end{document}